\documentclass[12pt]{iopart}

\usepackage{graphicx}
\usepackage{hyperref}
\usepackage{textcomp}

\newcommand{\equationname}[1]{\textsc{Eq.}~#1}

\usepackage{color}

\begin{document}

\title{A quantum relay chip based on telecommunication integrated optics technology}

\author{A. Martin, O. Alibart, M. P. De Micheli, D. B. Ostrowsky, and S. Tanzilli}

\address{Laboratoire de Physique de la Mati\`ere Condens\'ee, CNRS UMR 7336, Universit\'e de Nice - Sophia Antipolis, Parc Valrose, 06108 Nice Cedex 2, France}
\ead{sebastien.tanzilli@unice.fr}

\begin{abstract}
We investigate an integrated optical circuit on lithium niobate designed to implement a teleportation-based quantum relay scheme for one-way quantum communication at a telecom wavelength. Such an advanced quantum circuit merges for the first time, both optical-optical and electro-optical non-linear functions necessary to implement the desired on-chip single qubit teleportation. On one hand, spontaneous parametric down-conversion is used to produce entangled photon-pairs. On the other hand, we take advantage of two photon routers, consisting of electro-optically controllable couplers, to separate the paired photons and to perform a Bell state measurement, respectively. 
After having validated all the individual functions in the classical regime, we have performed a Hong-Ou-Mandel (HOM) experiment to mimic a one-way quantum communication link. Such a quantum effect, seen as a prerequisite towards achieving teleportation, has been obtained, at one of the routers, when the chip was coupled to an external single photon source.
The two-photon interference pattern shows a net visibility of 80\%, which validates the proof of principle of a ``quantum relay circuit'' for qubits carried by telecom photons. In case of optimized losses, such a chip could increase the maximal achievable distance of one-way quantum key distribution links by a factor 1.8. Our approach and results emphasize the high potential of integrated optics on lithium niobate as a key technology for future reconfigurable quantum information manipulation.
\end{abstract}


\maketitle
\tableofcontents

\section{Introduction}

The field of quantum information relies on the attractive possibility of exploiting the quantum realm to accomplish tasks not accessible to traditional information processing and communication systems~\cite{Nielsen00}. This prospect has, over the two last decades, attracted wide interest among scientists from a wide range of disciplines including physics, mathematics, information theory, and engineering~\cite{Zoller_05}. Quantum superpositions of states and en\-tan\-glement are now seen as resources to treat quantum bits (qubits) of information~\cite{weihs_photonic_2001}. From the fundamental side, entanglement permitted, for instance, testing non-locality over long distances~\cite{Marcikic_50km_04,Ursin_entanglement_2007,Hubel_100km_07,Dynes_200km_09}. From the more applied side, quantum key distribution offers a secure means to establish secret keys between distant partners, useful for one-time-pad encryption, with a security level unattainable with classical methods~\cite{gisin_QKD_2002,duvsek_quantum_2006}. Despite encouraging results, no actual breakthrough, in terms of maximal distance separating the partners, has been achieved so far. Basically, inherent losses in fibre channels and dark-counts in the detectors prevent extending the distance to more than 200\,km~\cite{takesue_07}. A viable possibility for extending the distance is based on quantum relay schemes, where the basic idea is to break the overall channel into several shorter sub-sections~\cite{Collins_QRelays_2005}. Here another quantum communication protocol, known as quantum teleportation, lies at the heart of quantum relays~\cite{bennett_teleporting_1993}.

While these are seminal experiments, it is only through a technology, such as integrated optics, that one can progress towards practical, standardized, low cost, interconnectable, and reconfigurable elements. Similarly to the extraordinary technological developments achieved in fibre optical telecommunications, integrated optics has proven to be a powerful enabling technology for realizing guided-wave optical quantum devices~\cite{Tanz_LPR_2011}, such as high efficiency entanglement sources~\cite{tanzilli_ppln_2002,martin_polar_2010} as well as elementary quantum functions, such as quantum interfaces~\cite{Tanz_Interface_2005} and memories~\cite{Saglamyurek_BroadbandWQM_2011}. Today's integration efforts are pushing one step further towards merging, onto the same chip, several elementary functions so as to achieve complex operations. For instance, Shor's quantum factoring~\cite{Politi09} and two-photon quantum random walks~\cite{Owens_QRW_11} have been demonstrated using dedicated integrated quantum optical circuits. In the framework of long-distance quantum communication, we show in this paper how integrated optics on lithium niobate permits realizing a telecom-like quantum relay chip that could achieve the quantum relay function, in a compact, stable, efficient, and user-friendly fashion. Such a novel quantum relay circuit merges, for the first time, both nonlinear electro-optical and optical-optical effects, in view of offering a reliable way to achieve long distance quantum communication. Furthermore, as will be discussed in the following, integrated optics on lithium niobate offers the possibility to implement electro-optically controllable functions, such as photon routers and/or (de-)multiplexers. This paves the way for the design of next generation, \textit{i.e.} showing reconfigurable capabilities, quantum optical circuits.

We start by describing our optical device with which on-chip teleportation could be realized and briefly discuss about the extension of the maximum achievable distance in one-way quantum communication. Then, we present the classical characterizations of the two implemented nonlinear effects. Finally, we demonstrate the proof-of-principle of the quantum relay function, based on the so-called Hong-Ou-Mandel (HOM) two-photon interference effect which lies at the heart of the teleportation protocol~\cite{HOM_dip_1987,DeRied_QRelay_2004,Halder_2times25km_2005,Aboussouan_HOM_10}. 

\section{A quantum relay chip based on integrated optics}
\label{Sec_principle}

\begin{figure}
\begin{center}
\includegraphics[width=\columnwidth]{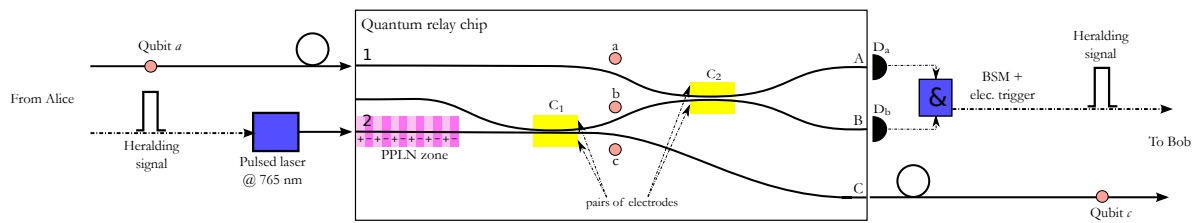}
\caption{Schematics of the quantum relay chip with its two electro-optically controllable couplers C$_1$ et C$_2$. D$_a$ and D$_b$ are two detectors responsible for the Bell state measurement (BSM). At the end of the quantum channel, Bob's detector is triggered by the AND-gate (\&) placed after these two detectors. The chip operation fits perfectly with the point-to-point telecommunication approach for which a classical heralding signal propagates through the channel along with the qubit.}
\label{IQR_chip}
\end{center}
\end{figure}

Figure{\,\ref{IQR_chip}} depicts the chosen integrated optical design for merging all the necessary functions on the same chip for enabling the process of teleportation which lies at the heart of the quantum relay operation~\cite{Collins_QRelays_2005,Aboussouan_HOM_10}. Let's suppose that Alice sends an unknown qubit $a$, for instance encoded on the time-bin observable~\cite{weihs_photonic_2001}, carried by a telecom wavelength photon, travelling along a fiber quantum channel connected to port $1$ of the relay chip. At the same time, a photon, from a pump laser pulse emitting in the visible range of wavelength, synchronized with the arrival time of qubit $a$, enters the nonlinear optical section of the chip (port $2$). This section consists of a channel waveguide integrated on periodically poled lithium niobate (PPLN). Thanks to an appropriate choice of the poling period, this pump photon can be converted, by spontaneous parametric down-conversion (SPDC), into a pair of, say time-bin, entangled photons ($b$ and $c$) whose wavelengths are identical to that of the photon carrying qubit $a$~\cite{tanzilli_ppln_2002}. Then, the first 50/50 directional coupler (C$_1$) is used to separate the created pairs in such a way that photons $a$ (sent by Alice) and $b$ enter the second 50/50 coupler (C$_2$) from each port.
Note that the afore mentioned synchronization requirement between $a$ and $b$ ensures that these photons enter C$_2$ simultaneously within their coherence time. Here, C$_2$ plays the role of a Bell state measurement (BSM) apparatus, which makes the latter requirement a crucial feature for having this operation working properly.
In this case, if conditions on the polarization states, central wavelengths, and coherence times are met, qubits $a$ and $b$ can be projected onto one of the four entangled Bell states identified when two detectors placed at the output of the chip (D$_a$ and D$_b$ in Figure{\,\ref{IQR_chip}}) fire simultaneously and lead to a registered coincidence ($\&$)~\cite{DeRied_QRelay_2004}. As a result of this measurement, the qubit initially carried by photon $a$ has been teleported to photon $c$ that exits the chip at port $C$, without, theoretically, any loss in its quantum properties~\cite{bennett_teleporting_1993}. This is made possible since qubits $c$ and $b$ are initially entangled, meaning, from the quantum point of view, that the key resource of entanglement has been consumed during the process. In addition, note that the initial qubit has not been cloned since photons $a$ and $b$ have been annihilated. As a consequence, the resulting electrical trigger from the measurement is not only the signature of the presence of the initial carrier at the relay chip location, but also of the departure of a photon carrying the same qubit that remains unknown. This allows triggering Bob's detectors at the end of the channel, therefore increasing the overall channel signal to noise ratio, and consequently the communication distance.

\begin{figure}[h!]
\begin{center}
\includegraphics[width=0.8\columnwidth]{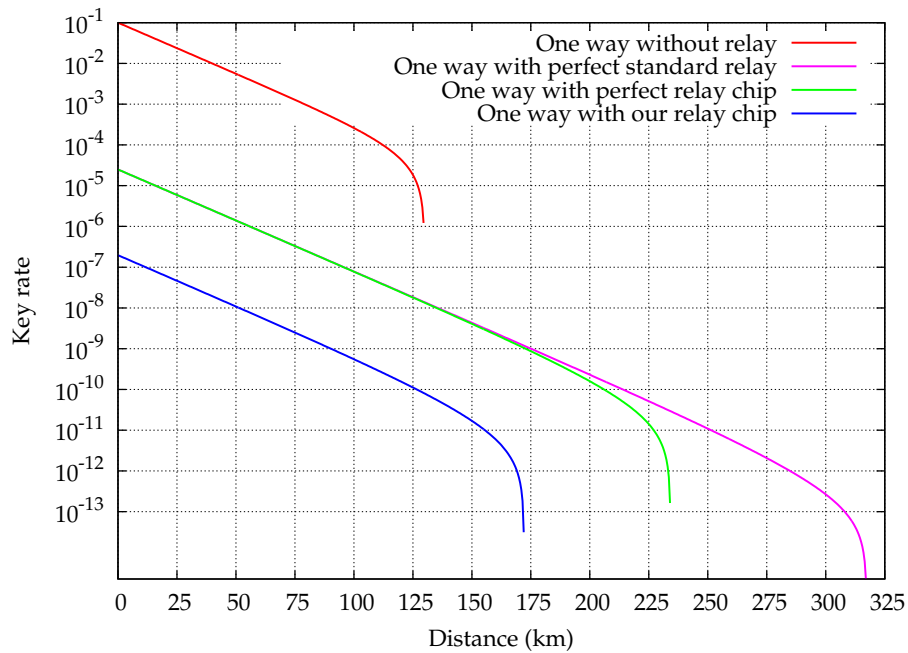}
\caption{Simulation of the normalized key rates as a function of the quantum communication distances for various one-way link configurations. Standard figures, used here, are fibre losses of $\sim$0.2\,dB/km, detectors' dark-count levels of $\sim$10$^{-6}$/ns associated with 10\% detection efficiencies (obtainable from commercially available InGaAs avalanche photodiode modules), and a quantum teleportation fidelity of 0.8 (as obtained in~\cite{DeRied_QRelay_2004}). This simulation proposes a comparison between a direct quantum communication link without any relay (red), and a standard quantum relay protocol based on the teleportation scheme (violet), together with its ``folded-version'' implemented using the actual relay chip (blue). In addition, the performance of a lossless relay chip is provided in order to show the potential improvement related to better technological fabrication of the chip (green). Absolute key rates can be obtained by multiplying the numbers on the y-axis by the pump laser repetition rate.}
\label{IQR_increase}
\end{center}
\end{figure}

This ``folded-version'' of the usual quantum relay is well adapted to the point-to-point telecommunication approach, \textit{i.e.}, a classical heralding signal propagating through the channel along with the teleported qubit. Moreover, it still improves the distance of any one-way quantum communication protocol and can obviously be more straightforwardly integrated within a telecom-like optical circuit. Figure{\,\ref{IQR_increase}} provides a comparison between normalized bit rates as a function of the distance for direct, as well as quantum relay based, one-way quantum communication links. On one hand, it shows that a chip, ideal in terms of both coupling and propagation losses, could improve the distance up to a factor of 1.8. On the other hand, a realistic chip, featuring loss figures as described in the following section, provides an increase of the distance by a factor of 1.4. In addition, these are compared with the usual lossless quantum relay scheme~\cite{Collins_QRelays_2005}.

\section{Realization and classical characterization}

From the technological side (see figure\,\ref{IQR_chip}), the chip features a waveguide quantum circuit consisting of a photon-pair creation zone and two tunable couplers merged onto a 5\,cm-long sample.

Generally speaking, the waveguiding structures are obtained by soft proton exchange (SPE)~\cite{Chanvillard_SPE_00}. This technique enables strong light confinement ($\delta n \simeq 2.2\cdot10^{-2}$) over long lengths, associated with both low propagation losses ($\sim$0.3\,dB/cm) and very high conversion efficiencies when nonlinear optical processes are implemented using PPLN sections~\cite{tanzilli_ppln_2002}. However, this technique leads only to an extraordinary index increase, meaning that only TM polarization modes can be guided in such structures. As a consequence, SPE waveguide devices cannot handle polarization qubits, as opposed to titanium in-diffused structures~\cite{martin_polar_2010,zhong_highquality_2010}.

\begin{figure}[h]
\begin{center}
\includegraphics[width=0.8\columnwidth]{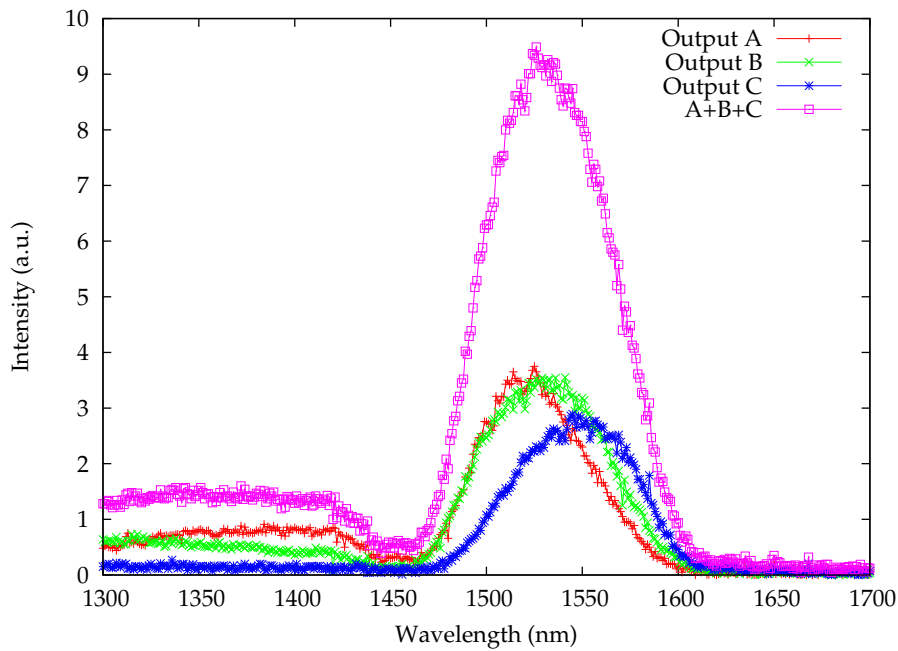}
\caption{SPDC spectra from the nonlinear waveguide zone. Outputs A and B correspond to the top outputs where the BSM is to be performed while the C output is the bottom output for the teleported qubit. Faithful characterization of the nonlinear interaction is provided by the sum of these three outputs. The difference between the three outputs is due to unwanted broadband wavelength sensitivity of the couplers. This behavior is explained by irregularities in the fabrication process leading to asymetric waveguides. Moreover, usual Cerenckov effect is observed on the left side of the fluorescence peaks~\cite{Ito_93}.}
\label{IQR_charac_SPDC}
\end{center}
\end{figure}

Considering the photon-pair generation section, the SPDC process is ruled by energy and momentum conservation laws. In this waveguide configuration, the latter is achieved using the so-called quasi-phase matching technique which compensates periodically for the dispersion between the three interacting photons (pump, $b$ and $c$). This allows phase-matching any desired wavelengths when the poling period is properly engineered. In our case, thanks to a poling period of 16.6\,$\mu$m, the photon-pair source is designed to ensure the conversion of 766\,nm pump photons into two degenerate photons at 1532\,nm (telecom C-band). Photon-pair sources based on SPE:PPLN waveguides have been demonstrated to be highly efficient and excellent providers of photonic entanglement~\cite{tanzilli_ppln_2002,Tanz_Interface_2005,halder_high_2008}. For our chip, we used a 1\,cm-long SPE:PPLN section which leads to the SPDC response shown in figure{\,\ref{IQR_charac_SPDC}} when pumped by a picosecond regime laser at 766\,nm and heated at a temperature of 80\,\textcelsius{}. One can see that the SPDC signal is distributed at the three output ports of the chip due to the presence of the two couplers. To characterize more completely the nonlinear interaction, the sum of these three contributions has been computed and added to the picture as the pink/square curve. We get here, as expected, degenerate paired photons around 1532\,nm, and the obtained spectral FWHM is of 80\,nm.

\begin{figure}
\begin{center}
\includegraphics[width=0.8\columnwidth]{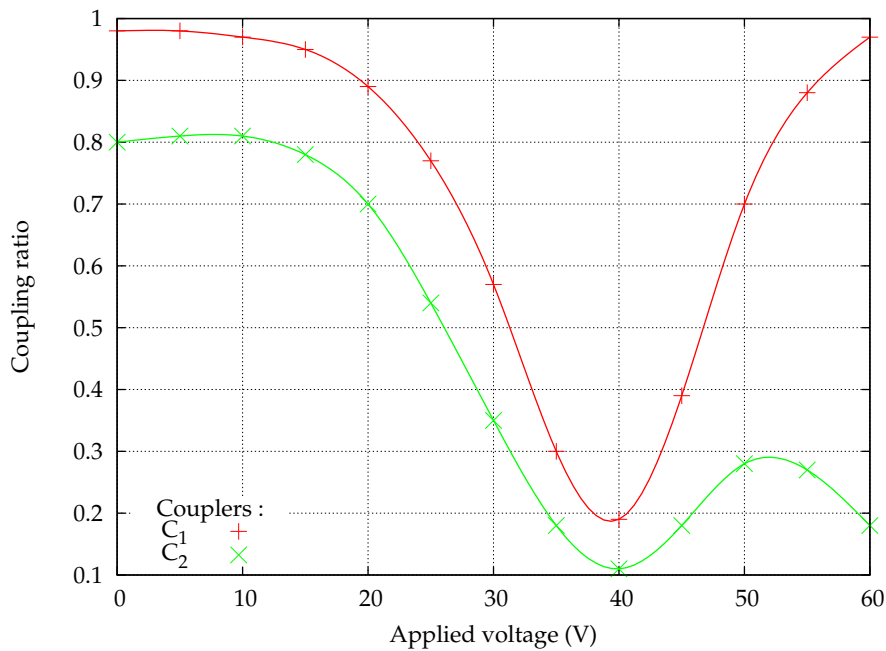}
\caption{Characterization of the coupling ratios obtained with couplers C$_1$ and C$_2$ in both narrowband classical light (lines) and broadband single photon light (crosses). The difference in coupling performance between C$_1$ and C$_2$ for voltages $<40$\,V is simply due to irregularities in the fabrication process.}
\label{IQR_charac_couplers}
\end{center}
\end{figure}

The directional couplers C$_1$ and C$_2$, represented in figure{\,\ref{IQR_chip}}, consist of two waveguides integrated close to each other over a given length ~\cite{Ostrow_75,Kogel76,wooten_review_2000}. If the spacing between the waveguides is sufficiently small, all the energy can be transferred from one to the other through evanescent coupling after a characteristic length. Afterwards, controllable decrease of the coupling ratio is achieved by detuning the propagation constant between the two waveguides by means of the nonlinear electro-optical effect. Numerical simulations have been carried out using the beam propagation method (BPM) in order to choose and optimize the bends, the spacing between the two waveguides, and the associated coupling length. Here the radius of the bends have to be carefully chosen since too small radii would introduce important losses while the opposite would lead to too long a chip. We selected 10\,mm long sinusoidal bends, corresponding to geometrical losses lower than 1\,dB regarding the propagation over the entire component. Note that the waveguides input/output spacing was imposed by external consideration since standard V-grooves with 250\,$\mu$m separated fibers were used at both end facets to couple photons in and out of the chip. With the constraint of single mode behaviour in the bends, we selected couplers made of 5\,$\mu$m-width waveguides separated by 4\,$\mu$m capable of ensuring a total energy transfer after a 9\,mm length. Electrodes, deposited on top of the two coupled waveguides with a silica buffer layer, are used to tune the coupling ratio from approximately 0:100 (transmission) to 100:0 (reflection). Figure{\,\ref{IQR_charac_couplers}} presents the characterization of the two couplers at 1532\,nm in the classical regime with a narrowband laser (Yenista Tunics) and in the photon counting regimes with a 1\,nm-bandwidth heralded single photon source at 1532\,nm, available in our lab~\cite{alibart_HSPS_2005}.

We see that the electro-optical control permits obtaining our necessary 50/50 coupling ratios for a voltage of around 30\,V. This is higher than the expected low-voltage ($<10$\,V) performance for such a configuration but still shows that, despite fabrication imperfections, the coupling ratio can be tuned continuously. The reason for this higher voltage value is essentially due to the important thickness of the silica buffer, deposited in-between the substrate and the metallic electrodes, used to reduce additional optical propagation losses due to metal loading. Moreover, note that slight misalignments between the electrodes and the couplers at the fabrication stage can also lead to increased control voltages. These results show that both obtained couplers fulfil the requirements, \textit{i.e.}, 50/50 ratios, necessary to implement on-chip quantum relay function.

Eventually, loss measurements at 1532\,nm over the entire chip length have been performed and shown to be on the order of 9\,dB between any couple of input to output fiber ports. These losses are in good agreement with the simulations we performed, taking into account input and output couplings ($2\times3$\,dB, for fiber-to-chip and conversely) and propagation losses over the entire chip (2.5\,dB). Note that integrating segmented taper waveguides at all input/output ports could help maximize the mode overlap between the fibers and the waveguides and therefore reduce the coupling losses by more than 1\,dB~\cite{castaldini_soft-proton-exchange_2007}.

\section{Proof-of-principle of the quantum relay function using the HOM effect}

Photon coalescence (or two-photon interference) based on the HOM effect lies at the heart of quantum operations and is seen as the major first step towards achieving teleportation~\cite{HOM_dip_1987,Landry2007}. Here, to demonstrate a proof of principle of a quantum relay function based on-chip BSM, we perform a HOM interference, or so-called HOM-dip, at the chip coupler C$_2$ between two independent photons, namely one external single photon and one photon out of the pair created on the chip.

To simulate the single photon coming from Alice's side, we use an external PPLN waveguide based source in the heralded single photon configuration, also producing degenerate paired photons at 1532\,nm when pumped at 766\,nm~\cite{Aboussouan_HOM_10}. We select, within the created SPDC bandwidth, pairs of photons using a set of two fiber Bragg grating (FBG) filters. Such filters reflect the bandwidth of interest, which is collected through additional fiber optical circulators. These are set to select energy-matched pairs with wavelengths around 1530.0\,nm ($\Delta\lambda=$200\,pm) and 1534.0\,nm ($\Delta\lambda=$800\,pm) on both sides of degeneracy. This filtering solution based on standard telecom components provides a clever way to separate the photons deterministically at the output of the PPLN waveguide source and makes them available in separate single mode fibers. The shortest wavelength photons are directed, after polarization adjustment, to the input port 1 of the chip, while the complementary ones are not used for that particular measurement for obtaining reasonable overall coincidence rates. These photons are only used to monitor the stability of the experiment. Basically, such a single photon source features the same photon statistics as an attenuated laser.

\begin{figure}
\begin{center}
\includegraphics[width=\columnwidth]{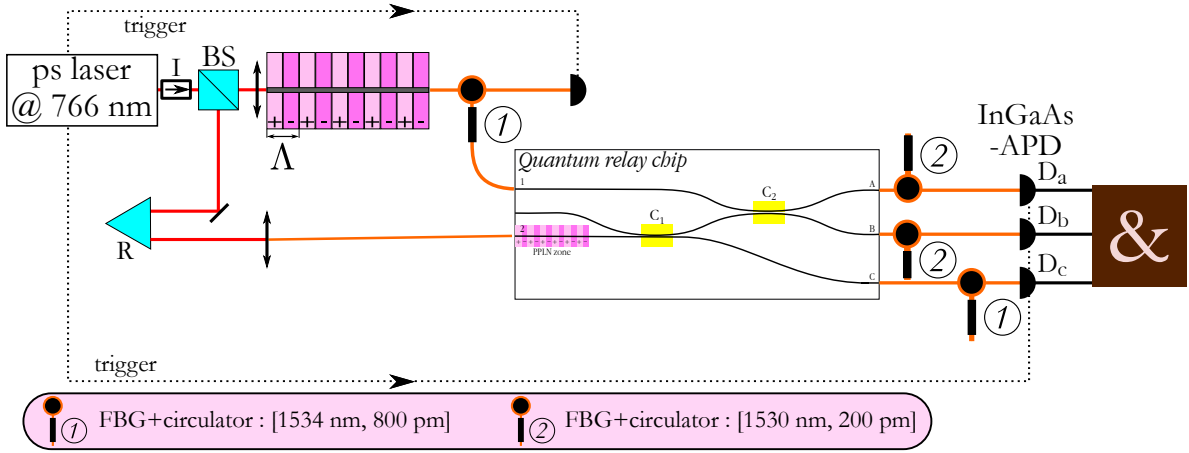}
\caption{Coalescence experiment based on one external PPLN waveguide based source and the quantum relay chip pumped by a single picoseconds laser. A silicon APD (not represented) is employed to trigger the four InGaAs-APDs as a laser clock random divider. D$_a$, D$_b$, and D$_c$ are used to record three-fold coincidences, while the non labelled detector is used to monitor the stability of the experiment. I: isolator; R: retro-reflector; C: circulator; \&: AND-gate. The InGaAs-APDs feature 10\% quantum efficiency and a dark-count probability of about 10$^{-5}$ ns$^{-1}$.
\label{fig_function_relay}}
\end{center}
\end{figure}

As depicted in figure{\,\ref{fig_function_relay}}, we use a picosecond pump laser (Coherent MIRA 900-D) emitting 2.5\,ps-duration, time-bandwidth limited ($\Delta \lambda_p$ = 250\,pm) pulses, at the wavelength of $\lambda_p$=766\,nm, and at a repetition rate of 76\,MHz. This laser is used to pump both the external PPLN waveguide source and the quantum relay chip to facilitate the synchronization procedure described in Sec.~\ref{Sec_principle}. The laser delivers mean power values of 1.5 and 7\,mW to the single photon source and the on-chip PPLN section, respectively. The photon-pairs ($b$ \& $c$) out of the on-chip PPLN section are randomly separated at the first coupler C$_1$. At the output of the chip, another set of filters, identical to the previous one, is used to post-select, on one hand, couples of photons $b$ and $a$ at 1530\,nm through output ports $A$ and $B$, and, on the other hand, the complementary photons $c$ at 1534\,nm through the port $C$ of the chip.

To observe the two-photon interference, or HOM-dip, at coupler C$_2$, three-fold coincidences are recorded using three InGaAs avalanche photo-diode (APD) modules (idQuantique-201) placed at the output of the chip. Since the three detectors are not able to handle trigger rates above 1\,MHz, they are triggered by an Si-APD placed in the path of an attenuated fraction of the pump beam (not represented in Figure~\ref{fig_function_relay}) to obtain an average triple gate opening at 600\,kHz, randomly selected from the 76\,MHz laser repetition rate.

When considering two-photon interference requirements, indistinguishability in terms of energy is ensured by the filtering stages, while that associated with polarization states is ensured by the properties of the waveguide structures themselves, since they support TM modes only. The expected HOM-dip visibility is nevertheless limited by the time uncertainty of the two photons entering coupler C$_2$, which reads~\cite{rarity_quantum_1997}:
\begin{equation}
 V_\mathrm{max}^\mathrm{uncert} = \frac{1}{\sqrt{\left(\frac{\tau_\mathrm{uncert}}{\tau_\mathrm{c}}\right)^2 + 1}}, 
\end{equation}
in which, considering our pulsed configuration, $\tau_\mathrm{uncert}$ and $\tau_\mathrm{c}$ represent the pulse duration of the pump laser and the interfering photons' coherence times, respectively.
Note that $\tau_\mathrm{uncert}$ takes into account any broadening of the pump. Here, in the picosecond regime, the contribution of the dispersion over the chip length to the actual pulse duration is negligible.
For this reason, the 1530\,nm filters, at the joint outputs $A$ and $B$, exhibit a bandwidth of 200\,pm, to reach photon coherence times of 17.3\,ps, corresponding to a maximal achievable visibility close to 100\%. Note that the 1534\,nm filters feature a 800\,pm bandwidth, \textit{i.e.}, four times larger than the afore mentioned 200\,pm bandwidth, to maximize the three-fold coincidence rate.

Moreover, the visibility of the dip also depends strongly on the statistics of the two single photons both entering one input of the coupler $C_2$. The visibility of the HOM-dip is therefore given by
\begin{equation}
V = \frac{P^c_{\rm max}-P^c_{\rm min}}{P^c_{\rm max}}\label{dip_vis},
\end{equation}
where $P^c_\textrm{max}$ and $P^c_\textrm{min}$ represent the probability of coincidence outside and inside the dip, respectively. For a low mean number of photons, these two probabilities are approximated by: 
\begin{equation}\label{Eq_pmin}
P^c_{\rm min} = P_{0,a} P_{2,b}+ P_{2,a} P_{0,b}
\end{equation}
and
\begin{equation}\label{Eq_pmax}
P^c_{\rm max} = P_{1,a} P_{1,b}+P_{0,a} P_{2,b}+ P_{2,a} P_{0,b},
\end{equation}
where $P_{n,j}$ represent the probability to have $n$ photons in the coupler input arm $j=a,b$. We can see that the maximal attainable visibility is limited by $P^c_{\rm min}$, or in other words, by multiple pair events arising from one source ($P_{2,j}$). The latter figure depends on the photon number distribution of the considered photon-pair generators.

\begin{figure}[h!]
\begin{center}
\includegraphics[width=0.8\columnwidth]{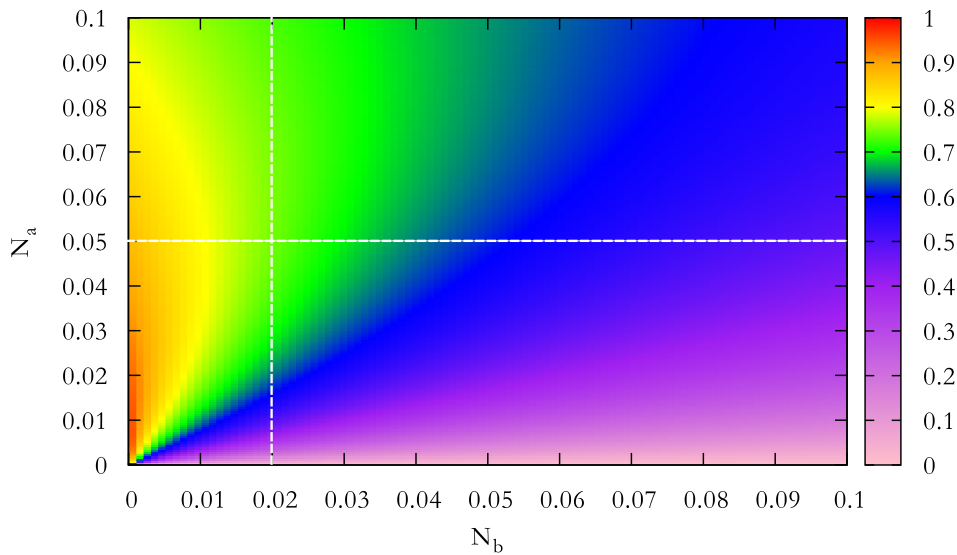}
\caption{\label{fig_Vmax}Maximal visibility as function of the average number of photons $N_b$ generated in the chip and the average photon number $N_a$ dispatched by Alice's source. Regarding the experimental setup, mean pump powers of 1.5 and 7\,mW in front of Alice's source and of the on-chip PPLN section lead to $N_a = 0.05$ and $N_b = 0.02$, respectively. The latter values allow reaching a theoretical visibility of 75\% (where the white dashed lines cross each other) while maintaining a reasonable coincidence rate.}
\end{center}
\end{figure}

In our experimental configuration, the paired photons are emitted by spontaneous parametric down-conversion. They are then filtered down to the time-bandwidth limit with respect to the pump pulse duration. The photon number distributions have been cha\-rac\-terized for each source, and both feature, as expected, thermal photon-pair distributions~\cite{tapster_photon_1998}. We define $N_a$ and $N_b$ as the average numbers of photon-pairs created per pump pulse out of Alice's and the chip generators, respectively. If we simply take into account two-fold coincidences between detectors D$_a$ and D$_b$, a theoretical visibility of 33\% can be reached when $N_a=N_b$, as described in~\cite{riedmatten_quantum_2003}. To improve the visibility figure, three-fold coincidences have to be measured between the three detectors placed at the output of chip. Consequently, the on-chip generator distribution follows that of a single photon source exhibiting a thermal statistics in the heralded regime~\cite{rarity_quantum_1997,pittman_violation_2003}. This permits reducing the probability $P_{0,b}$ associated with no photon-pair events (see \equationname{\ref{Eq_pmin}} and \ref{Eq_pmax}). 
Figure~\ref{fig_Vmax} presents the maximal achievable visibility obtained when solving \equationname{\ref{dip_vis}} as a function of $N_a$ and $N_b$ in the three-fold coincidence regime. In this case, mean pump powers of 1.5 and 7\,mW in front of Alice's and the chip generators, corresponding to $N_a = 0.05$ and $N_b = 0.02$, respectively, allow improving the theoretical visibility to 75\%. Finally note that reaching 100\% visibility is possible when $P_{0,a} \simeq P_{0,b} \simeq 0$, corresponding to the case where both photon sources are operated with thermal distributions and in the heralded regime~\cite{riedmatten_quantum_2003}.

\begin{figure}[h!]
\begin{center}
\includegraphics[width=0.8\columnwidth]{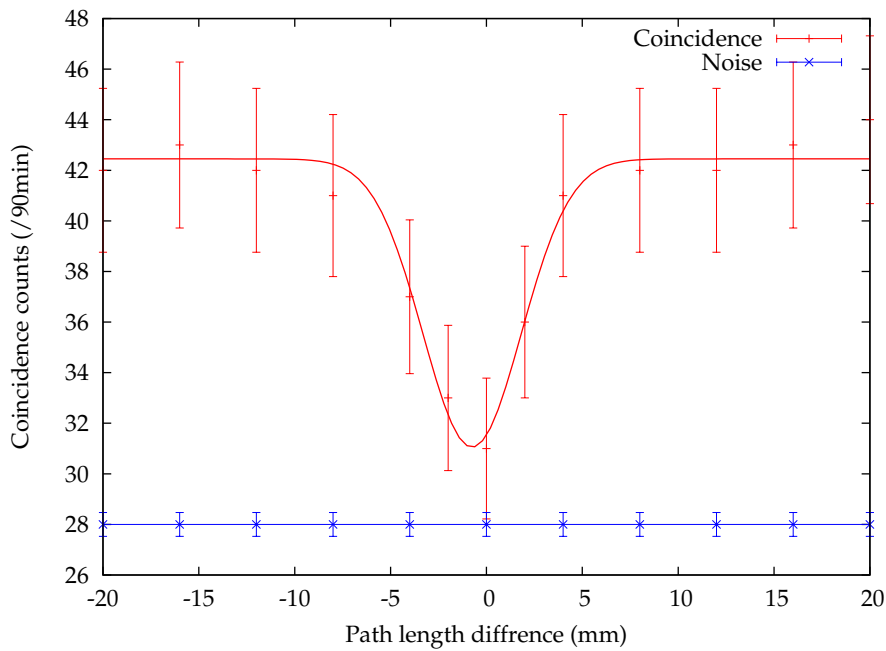}
\caption{\label{fig_dip}Three-fold coincidence rate as a function of the path length difference between the two interfering photons. The Gaussian fit of the interference pattern shows a raw (net) visibility of 27\%$\pm
$10\% (79\%$\pm$25\%).}
\end{center}
\end{figure}

Figure{\,\ref{fig_dip}} represents the evolution of the effective three-fold coincidence rate, measured as a function of the path length difference between the two interfering photons. The path length is adjusted using a retro-reflector ($R$) placed in front of the chip, as depicted in figure{\,\ref{fig_function_relay}}. When the two photons are made indistinguishable in terms of arrival times at the coupler C$_2$, a reduction of 27\%$\pm$10\% in the raw three-fold coincidence rate is observed. After subtraction of the accidental events due to both dark-count/dark-count and photon/dark-count contributions, we reach a net visibility of 79\%$\pm $25\%, which perfectly matches the above theoretical description and what we expect from figure{\,\ref{fig_Vmax}}. %As discussed earlier, the visibility of the dip is mainly limited by multiple pair events, which cannot be neglected with the chosen average numbers of pairs per pulse.
Furthermore, a full width at half maximum of 6\,mm is obtained, corresponding to a coherence time of about 20\,ps, which is in good agreement with the photons' coherence time of 17\,ps associated with the 200\,pm bandwidth filters. 

Both the large error bars and the low raw visibility are essentially due to a non-optimal detection scheme leading to a very low three-fold coincidence rate. These figures could be highly improved by the use of a new generation of InGaAs detector capable of being triggered at higher rates up to 100\,MHz. Compared to our current detection scheme that allows exploiting less then 1\% of the laser pulses, such modules would make it possible to attain a 2 order of magnitude improvement of the three-fold coincidence rate. This would enable the realization of the same experiment with a substantial reduction of both the error bars and the integration times. Eventually, this would allow taking in to account the heralding signal from Alice, enabling one to achieve the theoretical net visibility of 100\%.

\section{Conclusion \& perspective}

We have demonstrated that all the elements for implementing an integrated quantum relay circuit have been designed and merged onto a single chip. Each optical function has been tested separately, namely the photon-pair source and the photon routers. The on-chip source emits paired photons by SPDC within the telecom C-band, while the couplers show a 50/50 ratio when controlled using electro-optical means. Using such a chip, we also demonstrated a proof-of-principle experiment, based on the so-called HOM two-photon interference effect, which is seen as the major preliminary step towards achieving teleportation. We obtained in this quantum regime a HOM-dip at the second router of the chip featuring 79\% net visibility using two independent single photons, one external and one out the on-chip created pairs. This result, together with the width of the dip, is in good agreement with the theoretical description associated with our experimental configuration. With our non ideal chip, presenting rather high overall losses, the achievable quantum communication distance could be possibly augmented by a factor of 1.4, whereas an optimized device could lead to a factor of 1.8.

We believe that utilizing, for the first time, both nonlinear optical-optical and electro-optical effects provides a significant demonstration of the applicability of integrated optical technology on lithium niobate for quantum information treatment and applications.
Ongoing work now consists in improving the overall chip design to obtain much lower fiber-to-fiber losses regarding any input/output couple, which would permit realizing a true teleportation experiment. Future directions would lead to designing a wavelength demultiplexing module at C$_1$ location that would allows exploiting every created pairs in the nonlinear section.  On the other hand, a suitable adaptation of the deposited electrodes on the routers would enable designing novel circuits, featuring reconfigurability capabilities, offering new perspectives in the field. On-chip and on-demand generation of various photon-number states dedicated to quantum logic gates operation could be envisioned~\cite{Tanz_LPR_2011}.

\section*{Acknowledgement}

The authors would like to thank G. Bertocchi and S. Tascu for their help in the chip manufacturing, P. Baldi and F. Doutre for discussions, M. Al Khalfioui for the micro-soldering of the electrodes, and G. Sauder for his support in software development. The authors also acknowledge the CNRS, the University of Nice - Sophia Antipolis, the Agence Nationale de la Recherche (ANR) for the ``e-QUANET'' project (grant agreement ANR-09-BLAN-0333-01), the European ICT-2009.8.0 FET Open program for the``QUANTIP'' project (grant agreement 244026), the Conseil Regional PACA for the Volet Exploratoire project ``QUANET'', and the Fondation iXCore pour la Recherche, for financial support.

\end{document}